\documentclass[aps,prl,twocolumn,superscriptaddress]{revtex4-1}

\usepackage{graphicx}
\usepackage{bm}
\usepackage{rotating}
\usepackage{amssymb,amsmath}
\usepackage{multirow}
 
\begin{document}

\title{Orbital occupancies and the putative $j_{\mathrm{eff}}=1/2$ groundstate
in Ba$_2$IrO$_4$: a combined oxygen K edge XAS and RIXS study}

\author{M.~Moretti~Sala}
\email{marco.moretti@esrf.fr}
\affiliation{European Synchrotron Radiation Facility, BP 220, F-38043 Grenoble
Cedex, France}

\author{M.~Rossi}
\affiliation{European Synchrotron Radiation Facility, BP 220, F-38043 Grenoble
Cedex, France}

\author{S.~Boseggia}
\affiliation{London Centre for Nanotechnology and Department of Physics and
Astronomy, University College London, London WC1E 6BT, United Kingdom}
\affiliation{Diamond Light Source Ltd, Diamond House, Harwell Science and
Innovation Campus, Didcot, Oxfordshire OX11 0DE, United Kingdom}

\author{J.~Akimitsu}
\affiliation{Department of Physics and Mathematics, Aoyama Gakuin
University, 5-10-1 Fuchinobe, Chuo-ku, Sagamihara, Kanagawa 252-5258, Japan}

\author{N.~B.~Brookes}
\affiliation{European Synchrotron Radiation Facility, BP 220, F-38043 Grenoble
Cedex, France}

\author{M.~Isobe}
\affiliation{National Institute for Materials Science (NIMS), 1-1 Namiki,
Tsukuba, Ibaraki 305-0044, Japan}

\author{M.~Minola}
\affiliation{CNR-SPIN, CNISM and Dipartimento di Fisica, Politecnico di Milano,
Piazza Leonardo da Vinci 32, I-20133 Milano, Italy}

\author{H. Okabe}
\affiliation{National Institute for Materials Science (NIMS), 1-1 Namiki,
Tsukuba, Ibaraki 305-0044, Japan}

\author{H.~M.~R\o{}nnow}
\affiliation{Laboratory for Quantum Magnetism, \'{E}cole Polytechnique
F\'{e}d\'{e}rale de Lausanne (EPFL), CH-1015, Switzerland}

\author{L.~Simonelli}
\affiliation{European Synchrotron Radiation Facility, BP 220, F-38043 Grenoble
Cedex, France}

\author{D.~F.~McMorrow}
\affiliation{London Centre for Nanotechnology and Department of Physics and
Astronomy, University College London, London WC1E 6BT, United Kingdom}

\author{G.~Monaco}
\affiliation{Dipartimento di Fisica, Universit\`a di Trento, via Sommarive 14,
38123 Povo (TN), Italy}
\affiliation{European Synchrotron Radiation Facility, BP 220, F-38043 Grenoble
Cedex, France}

\begin{abstract}

The nature of the electronic groundstate of Ba$_2$IrO$_4$ has been 
addressed using soft X-ray absorption and inelastic scattering
techniques in the vicinity of the oxygen K edge. From the polarization and
angular dependence of XAS we deduce an approximately equal superposition of
$xy$, $yz$ and $zx$ Ir$^{4+}$ 5d orbitals. By combining the measured orbital
occupancies, with the value of the spin-orbit coupling provided by RIXS, we 
estimate the crystal field splitting associated with the tetragonal distortion
of the IrO$_6$ octahedra to be small, $\Delta\approx50(50)$ meV. We thus
conclude definitively that Ba$_2$IrO$_4$ is a close realization of a spin-orbit
Mott insulator with a $j_{\mathrm{eff}}=1/2$ groundstate, thereby overcoming
ambiguities in this assignment associated with the interpretation of X-ray
resonant scattering experiments.

\end{abstract}

\maketitle

The comprehensive characterisation of the groundstate of a system forms
the basis for understanding its physical properties. 
In the case of the new class of spin-orbit induced Mott insulating
iridium oxides, the so-called $j_{\mathrm{eff}}=1/2$ groundstate has been
proposed to explain the unexpected transport properties of Sr$_2$IrO$_4$ (and
related compounds)\cite{Moon2008,Kim2008} and is more generally central to the prediction
of the novel electronic phases in iridates.
%\cite{Jackeli2009,Shitade2009,Wang2011,XiangangWan2011,Chaloupka2013}. 
Naively one might predict that a 5$d$ transition-metal oxide (TMO) would be
metallic because of the large extension of its orbitals and reduced electronic
correlations. However, the large spin-orbit coupling present in the 5$d$
transition-metal leads to the lifting of degeneracies in the groundstate
manifold, allowing diminished electronic correlations to play an enhanced role.
Accordingly, Sr$_2$IrO$_4$ turns out in fact to be an antiferromagnetic
insulator with properties reminiscent of 3$d$ TMO Mott insulators
\cite{Crawford1994}. 

Valuable insight into the physics of Sr$_2$IrO$_4$ and related materials
can be obtained within the frame of a single-ion model, in which 5$d$ e$_g$
states are neglected and one \emph{hole} fills the t$_{2g}$ states, which are
subject to tetragonal crystal field splitting ($\Delta$) and spin-orbit
coupling ($\zeta$)\cite{Ament2011a,Liu2012,Ohgushi2013,Hozoi2012}. For
magnetic moments oriented along the (110) direction, the ground state wave
function is written as\cite{MorettiSala2013_arXiv}
\begin{equation}
 |0,-\rangle = \frac{C_0 \left( |xy,-\rangle
-\imath |xy,+\rangle\right)/\sqrt{2}+ |yz,+\rangle
+ \imath |zx,-\rangle}{\sqrt{2+C_0^2}}, \label{gs_wavefunction2}
\end{equation}
with $2C_0=1-\delta-\sqrt{9+\delta(\delta-2)}$ ($C_0\leq0$), where
$\delta=2\Delta/\zeta$. Theoretically, a pure $j_{\mathrm{eff}}=1/2$ ground
state is realized for $\Delta=0$ only, with an equal occupation of the $xy$,
$yz$ and $zx$ orbitals. Instead, for $\Delta\gg\zeta$ the $\left( |xy,-\rangle
-\imath |xy,+\rangle\right)/\sqrt{2}$ state is stabilized, while
for $\Delta\ll-\zeta$ the ground state reduces to $\left( |yz,+\rangle +
\imath|zx,-\rangle\right)/\sqrt{2}$.

Compelling, microscopic evidence for the existence of a 
$j_{\mathrm{eff}}=1/2$ ground state in iridium oxides was first provided by 
X-ray resonant magnetic scattering (XRMS) experiments.
It was argued that the near vanishing of the L$_2$/L$_3$ 
magnetic Bragg peak intensity ratio provides a unique signature of the
$j_{\mathrm{eff}}=1/2$ groundstate\cite{Kim2009,JWKim2012}. 
The subsequent application of XRMS to several compounds led to the identification of a 
number of other ``$j_{\mathrm{eff}}=1/2$ iridates'', such as
Sr$_3$Ir$_2$O$_7$\cite{Boseggia2012,Boseggia2012a,JWKim2012},
Ba$_2$IrO$_4$\cite{Boseggia2013}, CaIrO$_3$\cite{Ohgushi2013} and
Ca$_4$IrO$_6$\cite{Calder2013_arXiv}. However, early on
Chapon and Lovesey\cite{Chapon2011} raised doubts on this interpretation.
More recently, some of the present authors further clarified the situation by
showing that the Ir L$_2$ edge XRMS
cross-section is identically zero \emph{irrespective} of the tetragonal crystal
field splitting of t$_{2g}$ states when the magnetic moments
lie in the \emph{ab}-plane\cite{MorettiSala2013_arXiv}. Therefore, 
a zero L$_2$/L$_3$ XRMS intensity ratio is a necessary, but not sufficient
condition for the $j_{\mathrm{eff}}=1/2$ groundstate to be realized in iridate
perovskites with in-plane easy magnetic axis. Sr$_2$IrO$_4$\cite{Kim2009} and
Ba$_2$IrO$_4$\cite{Boseggia2013} fall exactly in this category.

By now, the $j_{\mathrm{eff}}=1/2$ ground state of Sr$_2$IrO$_4$ has been verified by a
number of experimental techniques, including optical conductivity, O K edge XAS,
ARPES\cite{Kim2008}, XRMS\cite{Kim2009} and
RIXS\cite{Ishii2011,Kim2012}, and the tetragonal crystal field has 
been shown to be small \cite{Boseggia2013a}. Like the benchmark compound, it was
suggested that Ba$_2$IrO$_4$ is a spin-orbit Mott
insulator\cite{Okabe2011}, and indeed recent ARPES
measurements suggest that Sr$_2$IrO$_4$ and Ba$_2$IrO$_4$ have qualitatively
similar band structures\cite{Moser2013_arxiv}. The crystal structure and the
magnetic properties of the two compounds, however, are significantly
different. In both Sr$_2$IrO$_4$ and Ba$_2$IrO$_4$ the apical
Ir-O bonds are longer than the in-plane bonds, but with differences of 4\%
and 7\%, respectively\cite{Crawford1994,Okabe2011}, implying a
larger tetragonal distortion in the latter compound.
Additionally, in Sr$_2$IrO$_4$ the IrO$_6$ octahedra have a staggered
rotation about the $c$ axis by $\sim12^\circ$\cite{Crawford1994}, such that the
in-plane Ir-O-Ir angles are not 180$^\circ$,
while in Ba$_2$IrO$_4$ the Ir-O-Ir bonds are straight\cite{Okabe2011}. The
combination of a larger tetragonal distortion and the absence of octahedra
rotation has mainly two effects: i) the crystal field potential acting on the
central Ir ion is different in the two compounds. \emph{Ab
initio} quantum chemistry calculations by L. Hozoi \emph{et
al.}\cite{Katukuri2012} show that the local Ir-O bonding sets a different energy
order of the Ir t$_{2g}$ levels for the two compounds in the absence of
spin-orbit coupling; in the \emph{electron} representation, the
lowest, doubly occupied t$_{2g}$ level is $xy$ in Ba$_2$IrO$_4$, and $yz/zx$ in
Sr$_2$IrO$_4$. ii) the inversion symmetry at the O site between adjacent
Ir ions is preserverd and Dzyaloshinsky-Moriya-type (DM) interaction is
suppressed. DM interaction in Sr$_2$IrO$_4$ is responsible for the canting of
the magnetic moments, which in turn produces a finite ferromagnetic component
below 240 K\cite{Kim2009}; Ba$_2$IrO$_4$, instead, does not exhibit a net
magnetic moment at any temperature\cite{Okabe2011}.

Thus taking into account all of the above facts, the question of whether or not 
the $j_{\mathrm{eff}}=1/2$ groundstate is realised in Ba$_2$IrO$_4$ is an interesting 
but open one. Here we address this question by presenting the results of 
oxygen K edge XAS and RIXS which allow us to directly determine the 
occupancy of the $xy$, $yz$ and $zx$ orbitals in the groundstate of Eq. 1., and to 
estimate the tetragonal crystal field.

\begin{figure}
	\centering
		\includegraphics[width=.99\columnwidth]{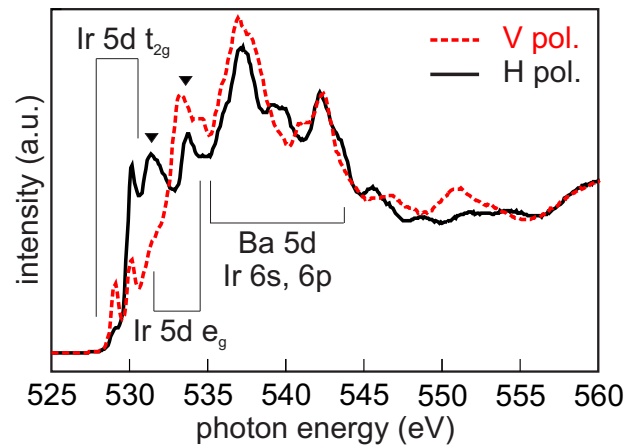}
\caption{XAS spectra of Ba$_2$IrO$_4$ for both
horizontal (black continuous) and vertical (red dashed line) polarization of the
photons at
$\theta=80^\circ$.}\label{fig1}
\end{figure}

O K edge x-ray absorption spectroscopy (XAS) and resonant inelastic x-ray
scattering (RIXS) data were taken at the ID08 soft x-ray beam line of the
European Synchrotron Radiation Facility, Grenoble. XAS spectra were
collected in total-fluorescence-yield mode to ensure the highest bulk sensitivity.
RIXS spectra were measured with the AXES spectrometer working at a fixed
scattering angle of 130$^\circ$. The energy-resolution of the incident photons
was better than 100 meV, while the overall energy-resolution of the RIXS
spectra was 180 meV, as determined by measuring the off-resonant elastic
scattering of amorphous graphite. The photon polarization was set to be either
horizontal (H) or perpendicular (V) to the scattering plane. The photons
impinge onto the sample at an angle $\theta$ with respect to the normal of the
sample surface. RIXS measurements were performed at $\theta=25^{\circ}$ (specular
geometry); in this geometry, spectra taken with horizontal and vertical
polarization (4 hours each) were found to be similar, and therefore were summed
up to reduce the error bars. The temperature of the sample was kept below 50 K
during the whole experiment. The sample, a single crystal of Ba$_2$IrO$_4$ (of
$\sim 0.5\times0.5\times0.2$ mm$^3$ size), synthesized at the National Institute
for Materials Science (NIMS) by the slow cooling technique under
pressure\cite{Okabe2011}, was oriented having the (010) and (001)
crystallographic direction in the horizontal plane. Note that in this
geometry the vertical polarization is always parallel to the (100) axis, while
the direction of the horizontal polarization relative to the sample changes with
$\theta$ according to $\boldsymbol\epsilon_H=(0,\cos\theta,\sin\theta)$.

\begin{figure}
	\centering
		\includegraphics[width=\columnwidth]{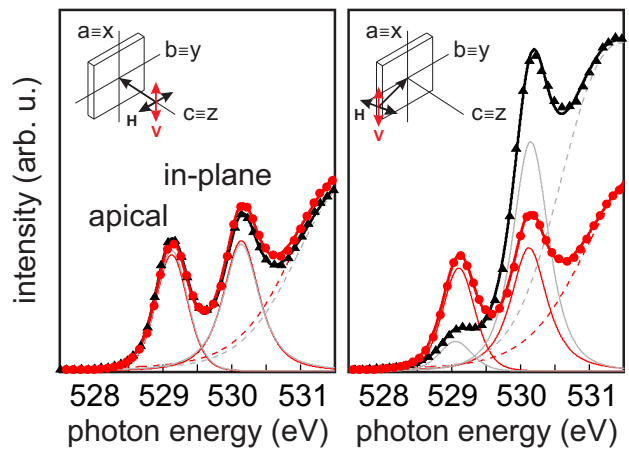}
\caption{XAS spectra of Ba$_2$IrO$_4$ for both
horizontal (black triangles) and vertical (red circles)
polarization of the photons at (left) $\theta=0^\circ$ and
(right panel) $\theta=80^\circ$. The corresponding experimental geometries are
shown in the inset. }\label{fig2}
\end{figure}

Fig.~\ref{fig1} shows two representative XAS spectra of Ba$_2$IrO$_4$ for both
horizontal (black) and vertical (red line) incident photon polarization at
$\theta=80^\circ$, over an extended energy range from 525 to 560 eV. A linear
background was subtracted by fitting the flat region at energies lower than 528
eV; the spectra were then normalized to unity on the high energy side of the
edge jump. We clearly observe a pronounced polarization dependence, most
notably in the pre-peak region, at energies lower than 535 eV. Typically,
this region is associated with transitions of the O 1$s$ core electrons to the
2$p$ states hybridized with the transition metal ion unoccupied valence states;
in our case, these can be mainly associated with the Ir 5$d$ t$_{2g}$ and e$_g$
states. Structures at higher energies are assigned to Ba 5$d$ and Ir
6$s$ and 6$p$ empty states. In Fig.~\ref{fig2}, we highlight the
polarization and angular dependence of the pre-peak region (528 to 533 eV)
by  focussing on the two peaks at 529.1 and 530.1 eV which arise from O $2p$-Ir
5$d$ t$_{2g}$ hybridization. At $\theta=0$ (left panel) the spectra
taken with the two photon polarizations are essential identical, while for
$\theta=80^\circ$ (right panel) the spectra are strongly affected by
polarization: specifically, the low-energy peak (529.1 eV) is
strongly suppressed, while the peak at higher energy (530.1 eV) is strongly
enhanced when switching from H to V polarization. This behaviour reflects the
anisotropy of the orbitals: at $\theta=0$, $\boldsymbol\epsilon_V=(100)$ and
$\boldsymbol\epsilon_H=(010)$, and the spectra are identical owing to the
equivalence of the (100) and (010) directions in Ba$_2$IrO$_4$; on the other
hand, $\boldsymbol\epsilon_V=(100)$, while $\boldsymbol\epsilon_H\approx(001)$
at $\theta=80^\circ$, and the spectra look different as Ba$_2$IrO$_4$ has
tetragonal symmetry.

Following the work of Moon \emph{et al.}\cite{Moon2006} on Sr$_2$IrO$_4$,
and in agreement with the peak assignments in
La$_{2-x}$Sr$_x$CuO$_4$\cite{CTChen1991}, Sr$_2$RuO$_4$\cite{Schmidt1996}, and
Ca$_2$RuO$_4$\cite{Mizokawa2001}, the two peaks at 529.1 and 530.1 eV of
Fig.~\ref{fig2} arise from transitions to the 5d t$_{2g}$ states at the apical
and in-plane oxygens, respectively. Also, the broad peaks at 532 and 534 eV
(indicated by triangles in Fig.~\ref{fig1}) correspond to transitions to the to
the 5d $3z^2$-$r^2$ and $x^2$-$y^2$ states, respectively\cite{Moon2006}. With
this assignments, 10Dq can be estimated to be about 3 eV in Ba$_2$IrO$_4$.

To quantitatively estimate their polarization and angular dependence, the XAS
spectra are fitted in Fig.~\ref{fig2} to two gaussian functions plus a third
curve (dashed lines) to account for higher-energy features. While the position
and full-width-half-maximum ($\approx0.55$ eV) of the peaks are insensitive to
the different geometries, it is interesting to look at the angular and
polarization dependence of the corresponding spectral weights: the integrated
intensity ratio of the apical and in-plane oxygen contributions is 0.71 (0.69)
for vertical (horizontal) polarization at $\theta=0$, and 0.64 (0.11) at
$\theta=80^\circ$, i.e. the apical oxygen contribution is strongly suppressed,
although not zeroed, in grazing incidence geometry and horizontal polarization.

\begin{figure}
	\centering
		\includegraphics[width=\columnwidth]{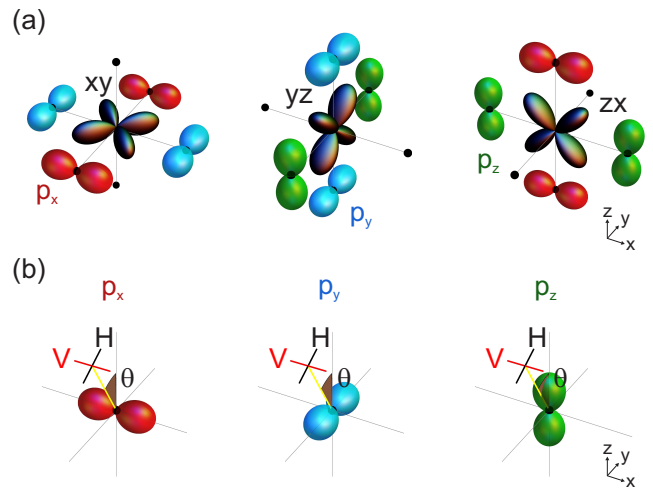}
\caption{(a) Sketch of the O 2$p_i$-Ir 5$d$ t$_{2g}$ orbital dependent
hybridizations, and (b) of the different O 2$p_i$ orbitals ($i=x,y$ and $z$)
orientations with respect to the incoming photon polarization. The
incidence angle $\theta$ is also indicated.}\label{fig3}
\end{figure}

The polarization and angular dependence of O K edge XAS is determined by the
O 2$p$-Ir 5$d$ t$_{2g}$ hybridization strength, and the dipole matrix element
governing the O 1$s$-2$p_i$ transitions ($i=x,y$ and $z$). Fig.~\ref{fig3}(a)
shows the O 2$p_i$-Ir 5$d$ t$_{2g}$ orbital dependent hybridizations; the $xy$
orbital has a finite overlap with 4 in-plane O 2$p_x$ and 2$p_y$ orbitals,
while the $yz$ ($zx$) orbital hybridizes with 2 apical O 2$p_y$ (2$p_x$) and 2
in-plane O 2$p_z$ (2$p_z$) orbitals. The strength of the hybridization, however,
depends on the Ir-O bond length $r$, which is different for the in-plane and
apical oxygens: here we assume a power-law dependence of the hybridization
strength on $r$, i.e. $r^{-\alpha}$, where $\alpha=3.5$\cite{Harrison1989}.
From Fig.~\ref{fig3}(b), one can understand the angular and polarization
dependence of the O 1$s$-2$p_i$ matrix elements, which, because of the isotropy
of the O 1$s$ orbital, is fully determined by the relative orientation of the
polarization vector and the O 2$p_i$ orbitals. It follows that the transitions
to the O $2p_x$, $2p_y$, and $2p_z$ orbitals have $\theta$ dependences of 0,
$\cos^2\theta$ and $\sin^2\theta$ for horizontal polarization, and 1, 0 and 0
for vertical polarization, respectively. Therefore, the intensity of the feature
associated with the in-plane oxygens has a $\theta$ dependence of
$n_{xy}\cos^2\theta +2 n_{yz}\sin^2\theta$ for horizontal polarization, and
$n_{xy}$ for vertical polarization. Here, $n_{xy}$ and $n_{yz}=n_{zx}$ are the
orbital occupancies of the $xy$ and $yz$ or $zx$ states, respectively
($n_{xy}+n_{yz}+n_{zx}=1$ in the case of Ir$^{4+}$). On the other hand, the
intensity of the feature associated with the apical component is given by
$1.07^{-\alpha}n_{yz}\cos^2\theta$ for horizontal polarization, and
$1.07^{-\alpha}n_{yz}$ for vertical polarization, where the factor
$1.07^{-\alpha}\approx80\%$ comes from the 7\% elongation of the IrO$_6$
octahedra. To summarize these arguments, no angular dependence is expected in
the case of vertical polarization, in agreement with the experimental findings,
while the intensity of the in-plane and apical oxygens contributions are
proportional to $n_{xy}$ and $n_{yz}$, respectively. In the case of horizontal
polarization, instead, an angular dependence is expected: specifically, at
normal incidence ($\theta=0$) the intensity is identical to the vertical
polarization case, while at $\theta=80^\circ$ the in-plane and apical oxygens
contribution are proportional to $0.03n_{xy}+0.97(2 n_{yz})$ and
$1.07^{-\alpha}0.03n_{yz}$, respectively. These considerations allow us to
extract the occupancies of the $xy$, $yz$ and $zx$ orbitals, in terms of
(fraction of) \emph{holes}, from the polarization and angular dependence of O K
edge XAS (Fig.~\ref{fig2}). We obtain $n_{xy}=0.37\pm0.04$ and
$n_{yz}=n_{zx}=0.31\pm0.02$. Although there appears to be a slightly higher
occupancy of the $xy$ orbital ($\Delta\gtrsim0$), this result is also consistent
with an almost even occupation of the $xy$, $yz$ and $zx$ orbitals
($n_{xy}=n_{yz}=n_{zx}=1/3$), suggesting that Ba$_2$IrO$_4$ belongs to the class
of $j_{\mathrm{eff}}=1/2$ iridates. 

\begin{figure}
	\centering
		\includegraphics[width=\columnwidth]{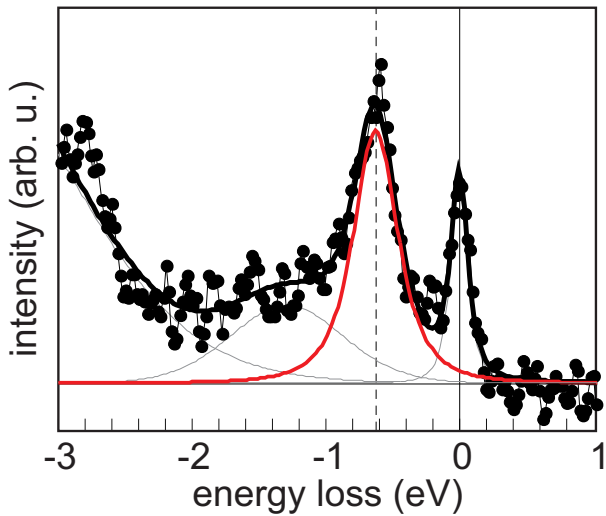}
\caption{O K edge RIXS spectrum of Ba$_2$IrO$_4$ for photons impinging at
the sample with an angle $\theta=25^\circ$ (red open circles). Solid lines are
the corresponding fitting curves used to determine the peak position.
}\label{fig4}
\end{figure}

Knowledge of the orbital occupancies allow us in turn to estimate the
\emph{effective} tetragonal crystal field splitting $\Delta$ (see Eq.\ 1 and 
related text). However, for this we first require a precise determination
of the spin-orbit coupling, $\zeta$. We can then use the fact the occupancies of
the $xy$, $yz$ and $zx$ orbitals are given by $n_{xy}=C_0^2/(C_0^2+2)$,
$n_{yz}=n_{zx}=1/(C_0^2+2)$, respectively, where the dependence on $\Delta$ and
$\zeta$ is parametrized in $C_0$. An estimate of the spin-orbit coupling
constant can be extracted from O K edge RIXS data. Fig.~\ref{fig4} shows a
representative O K edge RIXS spectrum of Ba$_2$IrO$_4$ for photons impinging on
the sample with an angle $\theta=25^\circ$ (specular geometry). The
spectrum shows a clear elastic line at zero energy loss and the first distinct
feature is found at 0.64 eV. Higher energy features have the characteristics of
resonant fluorescence emission, as verified by inspecting their incident photon
energy dependence. The incident photon energy was fixed at 530.1 eV, i.e.
the energy corresponding to the in-plane oxygen contribution in the O K edge XAS
spectra; doing so, we expected to maximize the intensity of RIXS transitions
involving orbitals hybridized with the Ir 5$d$ t$_{2g}$ states. Therefore, in
analogy to Ir L$_3$ edge RIXS, we assign this transition to the excitation of
the hole from the $j_{\mathrm{eff}}=1/2$ ground state to the
$j_{\mathrm{eff}}=3/2$ band. In a single ion
picture\cite{Ament2011a,Liu2012,Ohgushi2013,Hozoi2012,MorettiSala2013_arXiv},
the energy cost for this transition is $3\zeta/2$, from which the
value of the spin-orbit coupling constant $\zeta$ is estimated to be 0.43 eV.
This value remarkably agrees with the estimate of the spin-orbit coupling in
Ba$_2$IrO$_4$ ($\zeta=0.42$ eV) given in Ref.~\onlinecite{Boseggia2013}, based
on the Ir L$_2$ and L$_3$ XAS branching ratio. 

\begin{figure}
	\centering
		\includegraphics[width=0.82\columnwidth]{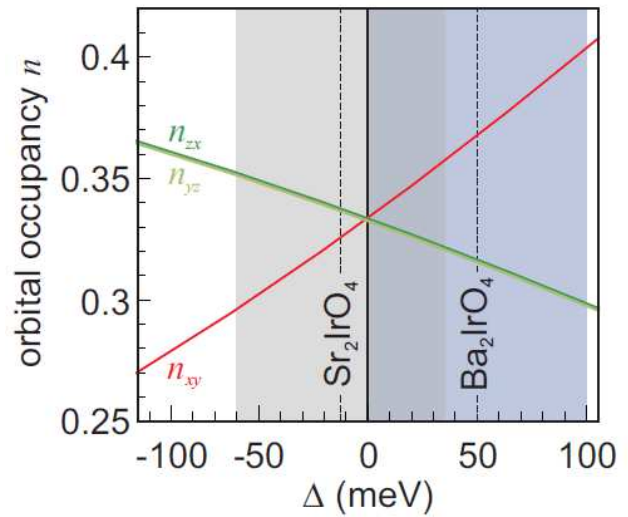}
\caption{Calculated orbital occupancies of the $xy$, $yz$ and $zx$ orbitals as
a function of the tetragonal crystal field splitting $\Delta$, for
$\zeta=0.43$ eV. The shaded areas indicate the estimated range of $\Delta$ in
Ba$_2$IrO$_4$, as extracted by the current experiment, and in Sr$_2$IrO$_4$
(Ref.~\onlinecite{Boseggia2013a}).}\label{fig5}
\end{figure}

Fig.~\ref{fig5} shows the calculated dependence of the orbital occupancies on
the tetragonal crystal field for $\zeta=0.43$ eV.
For the derived values of the orbital occupancies given above, we calculate
an \emph{effective} tetragonal crystal field of 50 meV, with
relatively tight constraints related to the error bars in the estimate of the
XAS peak intensities, i.e. $0\leq\Delta\leq100$ meV. This result may be compared
with the recent work of Boseggia \emph{et al.}\cite{Boseggia2013a} on
Sr$_2$IrO$_4$  who found that $-60\leq\Delta\leq35$ meV based on
the accurate determination of the magnetic moment direction relative to the
octahedra rotation by XRMS. Fig.~\ref{fig5} compares the
estimated tetragonal crystal field splittings in Sr$_2$IrO$_4$ and
Ba$_2$IrO$_4$. One can speculate that $\Delta\gtrsim0$ in Ba$_2$IrO$_4$, while
$\Delta\lesssim0$ in Sr$_2$IrO$_4$, i.e. the \emph{effective} tetragonal crystal
field has opposite sign in the two compounds. Although one can intuitively
explain this difference by invoking the elongation of the IrO$_6$ octahedra,
which is more pronounced in Ba$_2$IrO$_4$ than in Sr$_2$IrO$_4$, and thus
favours the occupancy of the $xy$ orbital, it should be noted that this result
contradicts recent theoretical calculations by Katukuri \emph{et
al.}\cite{Katukuri2012}.

In summary, we demonstrate how the combination of soft X-ray techniques yields
more information than any one technique used in isolation, allowing us in this
case to understand the link between structural distortions (octahedral rotation
and elongation) and the $j_{\mathrm{eff}}=1/2$ groundstate in iridates. 
By exploiting x-ray linear dichroism at the O K absorption edge of 
the prototypical perovskite iridate Ba$_2$IrO$_4$, we have determined the
orbital occupancies of Ir$^{4+}$ 5d groundstate to be $n_{xy}=0.37\pm0.04$ and
$n_{yz}=n_{zx}=0.31\pm0.02$. An estimate of the \emph{effective} tetragonal
distortion to the cubic crystal field splitting is given by extracting the
spin-orbit coupling constant ($\zeta=0.43$ eV) from O K edge RIXS measurements.
The result is that $\Delta=50(50)$ meV i.e. very small compared
with the cubic crystal field in Ba$_2$IrO$_4$ of about 3 eV. We conclude that
Ba$_2$IrO$_4$ may be classified as a spin-orbit Mott insulator, with a
groundstate close to the ideal $j_{\mathrm{eff}}=1/2$ state proposed to give
rise to the novel properties of iridates. We hope that the quantitative
information provided by our analysis will help refine future theoretical
approaches and act as a  guide to the large community of ``material engineers''
which exploits structural distortions to tune material properties.

\emph{Acknowledgments} - H.~M.~R\o{}nnow acknowledges support from the Swiss
National Science Foundation and its Sinergia network ``Mott Physics Beyond the
Heisenberg Model''. Work in London was supported by EPSRC.

%\bibliography{biblio}

\end{document}